\documentclass[alpha-refs]{wiley-article}

\usepackage{graphicx}
\usepackage[space]{grffile}
\usepackage{latexsym}
\usepackage{textcomp}
\usepackage{longtable}
\usepackage{tabulary}
\usepackage{booktabs,array,multirow}
\usepackage{amsfonts,amsmath,amssymb}
\usepackage{natbib}
\usepackage{url}
\usepackage{hyperref}
\hypersetup{colorlinks=false,pdfborder={0 0 0}}
\usepackage{etoolbox}
\newif\iflatexml\latexmlfalse

\AtBeginDocument{\DeclareGraphicsExtensions{.pdf,.PDF,.eps,.EPS,.png,.PNG,.tif,.TIF,.jpg,.JPG,.jpeg,.JPEG}}

\usepackage[utf8]{inputenc}
\usepackage[english]{babel}

\usepackage{siunitx}
\usepackage{overpic}
\usepackage{subfig}
\usepackage{caption}
\usepackage{amsfonts}

\iflatexml

\else

\corraddress{Fabian Mockert, Institute of Meteorology and Climate Research (IMK-TRO), Department Troposphere Research, Karlsruhe Institute of Technology (KIT),PO 3640, 76021 Karlsruhe, Germany}
\corremail{fabian.mockert@kit.edu}
\presentadd{Institute of Meteorology and Climate Research (IMK-TRO), Department Troposphere Research, Karlsruhe Institute of Technology (KIT), PO 3640, 
76021 Karlsruhe, Germany}
\fundinginfo{Helmholtz Association: VH-NG-1243 Young Investigator Group 'SPREADOUT' and Young Investigator Group VH-NG-1352, Deutsche Forschungsgemeinschaft: SFB/TRR 165 ‘Waves to Weather’, KIT Centre MathSEE}
\fi
\title{\vspace{-2cm}Meteorological conditions during Dunkelflauten in Germany: Characteristics, the role of weather regimes and impacts on demand}
\author[1]{Fabian Mockert}
\author[1]{Christian M. Grams}
\author[2,3]{Tom Brown}
\author[2,3]{Fabian Neumann}
\affil[1]{Institute of Meteorology and Climate Research (IMK-TRO), Department Troposphere Research, Karlsruhe Institute of Technology (KIT), Karlsruhe, Germany}
\affil[2]{Institute of Energy Technology, Department of Digital Transformation in Energy Systems, Technische Universität Berlin, Berlin, Germany}
\affil[3]{Institute for Automation and Applied Informatics (IAI), Karlsruhe Institute of Technology (KIT), Karlsruhe, Germany}
\date{\vspace{-0.5cm}December 8, 2022}
\begin{document}
\maketitle
\begin{abstract}
Renewable generation from wind and solar power is strongly weather-dependent.
To plan future sustainable energy systems that are robust to this variability, a better understanding of why and when periods of low wind and solar power output occur is valuable. We call such periods of low wind and solar power output `Dunkelflauten', the German word for dark wind lulls. In this article, we analyse the meteorological conditions during Dunkelflauten in Germany by applying the concept of weather regimes. Weather regimes are quasi-stationary, recurrent, and persistent large-scale circulation patterns which explain multi-day atmospheric variability (5-15 days). We use a regime definition that allows us to distinguish four different types of blocked regimes, characterised by high pressure situations in the North Atlantic-European region.
We find that in Germany, Dunkelflauten mainly occur in winter when the solar power output is anyway low and when the wind power output drops for several consecutive days. A high-pressure system over Germany, associated with the European Blocking regime, is responsible for most of the Dunkelflauten. Dunkelflauten during the Greenland Blocking regime are associated with colder temperatures than usual, causing higher electricity demand and presenting a particular challenge as space heating demand electrifies in future. Furthermore, we show that Dunkelflauten occur predominantly when a weather regime is well-established and persists longer than usual. Our study provides novel insight on the occurrence and meteorological characteristics of Dunkelflauten, which is essential for planning resilient energy systems and supporting grid operators to prepare for potential shortages in supply.\newline
\textbf{Keywords} Dunkelflauten, weather regimes, renewable energies, energy demand, predictability, Subseasonal prediction%
\end{abstract}

\section{Introduction}\label{Introduction}
One of the main objectives of the European Green Deal is to achieve climate neutrality for all countries in the European Union by 2050, in line with the Paris Agreement to limit global temperature increases to well below \SI{2}{\celsius} \citep{EU2019}. A key strategy to achieve climate neutrality is to raise the share of wind and solar power in Europe's electricity supply, and then to electrify as much of energy demand as possible. Wide-ranging electrification of all sectors would lead to a drastic increase in electricity consumption in Germany by 2045 \citep{TSO2022}. Depending on the scenario, German transmission system operators (TSOs) expect the gross electricity consumption to double by 2045 compared to 2020. To serve this demand, the TSOs project the installed capacities of onshore wind, offshore wind, and solar PV in Germany to rise from \SI{116}{\giga\watt} in 2020 up to \SI{616}{\giga\watt} in 2045 \citep{TSO2022}.\newline
With the increase in the share of variable renewable generation in the power system and the increase in electricity demand due to electrification \citep{IPCC2011, Bloomfield2021a}, the power system becomes increasingly sensitive to meteorological conditions \citep{Bloomfield2016, VanderWiel2019a}. Consequently, there is a rising need for balancing the spatial and temporal variability of renewable power output to guarantee a stable electricity supply. Seasonal variability can be balanced locally by the co-deployment of wind and solar technologies \citep{Pozo2004, Heide2010, Santos2015}. Daily variations of solar power can be balanced by storage or through flexibility from demand side management by the likes of battery electric vehicles \citep{Brown2018}. Power to gas units and long-term thermal energy storage were shown to support balancing large-scale and seasonal variations of electricity supply and demand. Some of these technologies can also be used to manage variability in power outputs over several days up to weeks. However, unlike for seasonal storage where the balancing needs are predictable, the operation on these multi-day time scales requires more detailed knowledge of the imminent meteorological conditions. Therefore, meteorological variability on a multi-day to weeks time range needs to be considered when planning a reliable energy system with a high share of renewable technologies \citep{Grams2017}.\newline
Meteorological conditions of multi-day events with low power output by renewable technologies can strain the energy system. In the following we refer to these periods of low solar and wind feed-in as `Dunkelflauten', from the German word for dark wind lulls.\newline
Dunkelflauten have been the subject of numerous studies under different names (low energy production and energy shortfall events \citep{VanderWiel2019a, VanderWiel2019b}, low production events (\citet{Kaspar2019} and extended by \citet{Drucke2021}), peak demand and peak demand-net-renewables events \citep{Bloomfield2020}, energy compound events \citep{Otero2022}). While \citet{VanderWiel2019a}, \citet{Bloomfield2020} and \citet{Otero2022} take the demand side into account, \citet{Kaspar2019} and \citet{Drucke2021} base their definition of Dunkelflauten exclusively on the availability of renewables.\newline 
Several studies (e.g. \citet{Drucke2021} for Germany and \citet{VanderWiel2019a} for Europe) suggest a high-pressure system in Central Europe is responsible for these low production events.
The large-scale meteorological conditions in the North Atlantic-European region during such events can be described by weather regimes. Weather regimes are quasi-stationary, persistent, and recurrent large-scale flow patterns in the midlatitudes \citep{Vautard1990, Michelangeli1995}. Weather regimes modulate surface weather in continent size regions and on multi-day to weekly time scales \citep{Bueler2021}. Thereby, weather regimes cause substantial multi-day variability on the European energy sector \citep{Zubiate2017,VanderWiel2019b} in particular wind power \citep{Grams2017}.\newline
To forecast Dunkelflauten events, or more generally the energy supply and demand, grid-point-based forecast methods or indirect pattern-based methods have been suggested \citep[cf.][]{Soret2019, Bloomfield2021b} . A grid-point forecast uses grid-point surface meteorological forecasts (e.g. \SI{10}{\meter} wind) to estimate a relevant power quantity (e.g. national wind power output). In a pattern-based forecast of the same power quantity, the large-scale atmospheric flow is first assigned to a preidentified circulation pattern, and in a second step, the surface impact is estimated. \citet{Bloomfield2021b} compare two pattern-based methods, the previously mentioned weather regime approach and an approach based on energy system data rather than large-scale meteorological fields, called targeted circulation types \citep{Bloomfield2020}. \citet{Bloomfield2021b} show that grid-point forecasts have higher skill at short lead times (days 0-10) than the pattern-based methods. At extended lead times (day 12+), pattern-based methods can show greater skill than the grid-point forecasts. Weather regime forecasts show higher skill in week\,3 (days 19-25) than targeted circulation types for Central and Northern European countries, likely owing to their physical grounding \citep{Faranda2016, Faranda2017, Hochman2021}. Forecasts on the time scale of 10-30\,days, also referred to as subseasonal to seasonal (S2S) forecasting range, become increasingly important for the energy sector as they fill the gap between weather forecasts and monthly or seasonal outlooks \citep{White2017, White2022}. We, therefore, link in our analysis Dunkelflauten to weather regimes.\newline
Blocking conditions are prone to cause below-average power output by wind and solar PV in Central Europe \citep{Grams2017, VanderWiel2019b}. The negative NAO phase is associated with cold and weak wind conditions \citep{Bloomfield2020, VanderWiel2019b, Tedesco2022}. However, not every blocking condition or negative NAO phase leads to Dunkelflauten in Germany. To accurately forecast Dunkelflauten events, more information about the weather regimes that lead to Dunkelflauten is necessary.\newline
In the present study, we aim to shed light on the meteorological conditions under which Dunkelflauten in Germany occur and how they are linked to the occurrence of weather regimes. Furthermore, we aim to understand how regimes related to Dunkelflauten differ from those not associated with a Dunkelflaute. \newline
This study is structured as follows. Data and methods are described in Section\,\ref{DataMethod}. In Section\,\ref{Results}, we describe the statistical characteristics of Dunkelflauten in Germany, discuss the meteorological conditions during Dunkelflauten and analyse the relationship between capacity factor, temperature and electricity demand anomalies during Dunkelflauten. Conclusions from this study and potential use cases of the presented results are highlighted in Section\,\ref{Discussion}.

\par\null

\section{Data and methods}\label{DataMethod}
\subsection{Reanalysis}\label{chap:reanalysis}
Reanalysis data from the European Centre for Medium-Range Weather Forecasts (ERA5) for 1979-2018 form the data basis for this study \citep{Hersbach2020}. The reanalyses serve as input data to calculate the capacity factors, compute the weather regimes and generate composites of atmospheric field variables during Dunkelflauten. We use ERA5 interpolated from its native reduced-Gaussian grid to a regular lat-lon grid with \SI{0.5}{\degree} grid spacing and a temporal resolution of three hours (\SI{0.25}{\degree} grid spacing and one hourly for the input in the energy system model, see below).

\par\null

\subsection{Weather regimes}\label{chap:weatherregimes}
To identify the large-scale atmospheric circulation during Dunkelflauten events, we use the definition of seven year-round weather regimes by \citet{Grams2017}. Regimes are identified based on an empirical orthogonal function analysis of 10\,day low--pass filtered geopotential height anomalies at \SI{500}{\hecto\pascal} (Z500) and a k--means clustering in the North Atlantic-European region(\SI{30}{\degree} to \SI{90}{\degree N}, \SI{80}{\degree W} to \SI{40}{\degree E}). Individual weather regime life cycles are identified using the projection of the instantaneous Z500 anomalous field into the cluster mean. The projection is normalised by the standard deviation over the entire ERA5 period to yield a weather regime index (IWR) for each of the 7 regimes following \citet{Michel2011}. During an active weather regime, the IWR is above 1.0 for at least 5 days and includes the maximum of all IWRs. The onset of the regime is defined as the first time step of $\textrm{IWR}>1.0$ \citep[cf.][]{Grams2017, Bueler2021}. Days not attributed to a regime life cycle are labelled as ``no regime'' days.\newline
The seven weather regimes can be separated into three cyclonic regimes (Atlantic Trough (AT), Zonal Regime (ZO) and Scandinavian Trough (ScTr)), which are characterised by low pressure systems in the European region, and four blocked regimes (Atlantic Ridge (AR), European Blocking (EuBL), Scandinavian Blocking (ScBL) and Greenland Blocking (GL)) which are dominated by high pressure systems.

\par\null

\subsection{Capacity factor}\label{chap:capacityfactora}
Our study focuses on the power output by renewable energy sources, namely solar, onshore and offshore wind power in Germany. As a measure for the power output, we use the resources' capacity factor time series, which denote the ratio between the actual power output and the rated capacity.\newline
The capacity factors are computed with the Python library atlite, which converts weather data (e.g. wind speed and solar influx) into energy system data (e.g. capacity factors) \citep{Hofmann2021}. To generate the capacity factor, atlite uses the ERA5 reanalysis dataset at \SI{0.25}{\degree} grid spacing and one hourly temporal resolution. For the wind capacity factor calculation, atlite uses the \SI{100}{\meter} wind speed and the surface roughness as input data. For calculating the solar capacity factor, the direct, diffuse and top of the atmosphere influx and the albedo are needed as input data.\newline
The geographic distribution of installed wind and solar capacities is assumed to be proportional to the technology's capacity factors (i.e. more wind farms in windy areas). We use a solar panel model based on \citet{Huld2010} with a solar azimuth of \SI{180}{\degree} (South) and a slope of \SI{35}{\degree}.
As a reference onshore wind turbine, we use the Vestas V112 3MW model, and for offshore the 5 MW NREL Reference Turbine.\newline
The total power output by renewable energy sources is expressed by combining the three different capacity factor time series to one combined capacity factor by applying a weighted mean with currently installed capacities per technology as weights. For Germany in 2018, these are: 44\% solar (45.9\,GW), 50\% onshore wind (53.0\,GW), 6\% offshore wind (6.4\,GW) \citep{IRENA2019}. Based on a linear regression between simulated and historical capacity factor time series, the combined capacity factors are corrected to match the observed time series \citep{OPSD2020} more closely.\newline 
The optimised approach of the geographic distribution of installed wind and solar capacities is likely to lead to a too high estimate of the capacity factors, but it is less dependent on political decision-making. The effect of the too-high capacity factor is damped by scaling the capacity factors with the actual power output of Germany given by \citet{OPSD2020}.
\begin{figure}[h!]
\includegraphics[width=1\columnwidth]{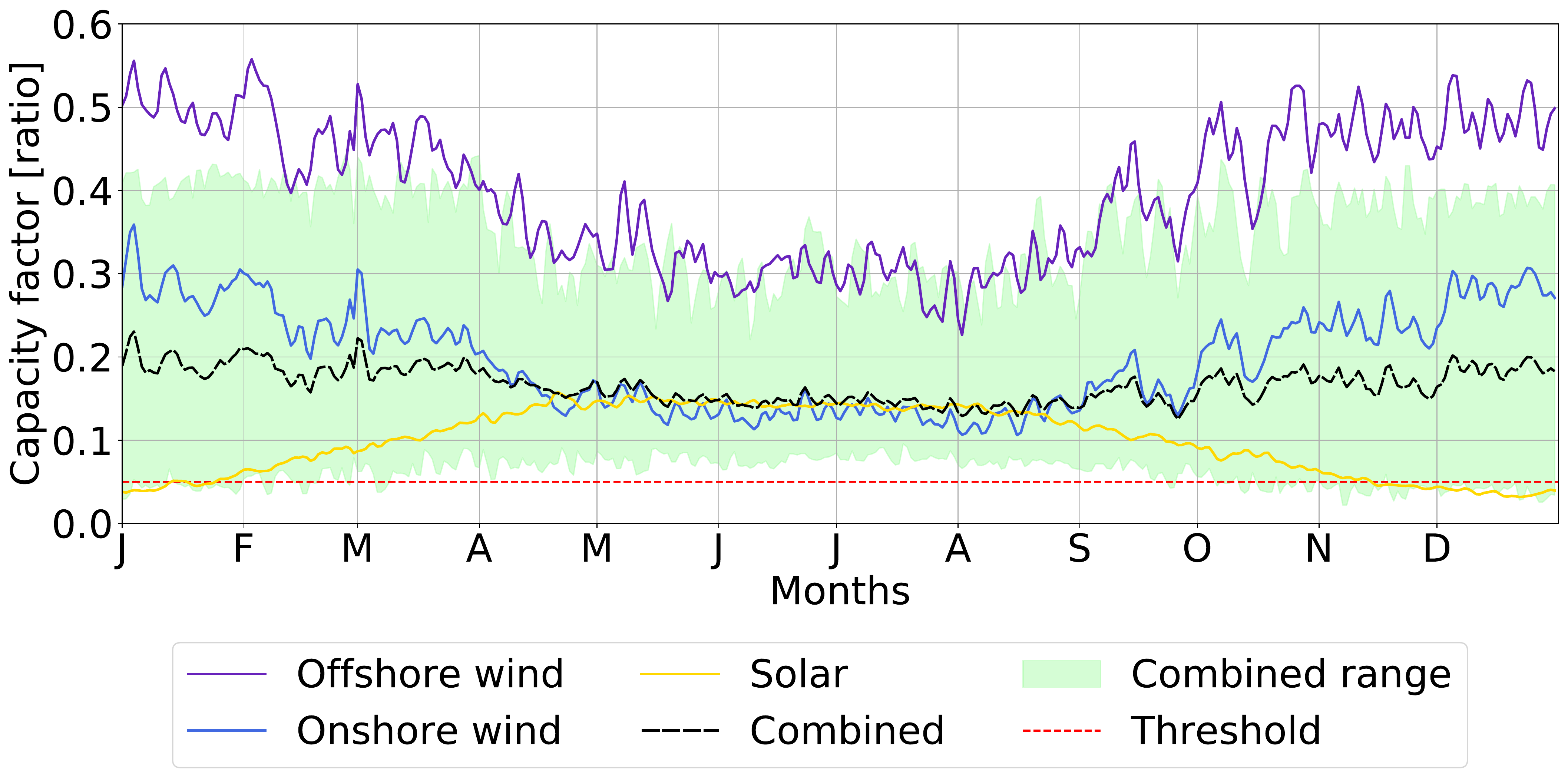}
\caption{Average seasonal cycle of the capacity factors in Germany for 1979-2018 (based on 3--hourly time steps as available for weather regimes). The combination of onshore wind (blue), offshore wind (purple) and solar (yellow) capacity factors, weighted on the installed capacity, is represented by the combined capacity factor (dashed black). The range (minimum, maximum) of the combined capacity factor throughout the 40\,year (1979--2018) period are visualised (mint green shading). The red horizontal line marks a threshold of 0.06.}
\label{fig:capacityfactor} 
\end{figure}\newline
For Germany, the individual capacity factors of wind and solar have a pronounced seasonal cycle (Figure\,\ref{fig:capacityfactor}). On average, the on- and offshore wind capacity factors are lower in summer than in winter, and the solar capacity factor is higher in summer than in winter. By combining the capacity factors of wind and solar, weighted by the installed capacity, the mean seasonal variation is balanced.\newline
From here on, we refer to the combined capacity factor when writing capacity factor, unless otherwise stated.

\par\null

\subsection{Dunkelflauten}\label{chap:dunkelflauten}
We are interested in periods when only little or no energy is available by wind and solar power for at least 2\,days. In this study, we define Dunkelflauten as periods with \SI{48}{\hour} running mean capacity factors below a threshold of 0.06. The frequency and length of Dunkelflauten events can be controlled by modifying the threshold value. Here, we set the threshold to 0.06 to obtain a similar number of Dunkelflauten per year as shown by \citet{Kaspar2019} (3--4\,Dunkelflauten per year with a threshold of 0.1). Our results regarding the central time of occurrence and meteorological characteristics of Dunkelflauten hold under modification of the threshold value (e.g. 0.05 or 0.10), although we detect less and shorter events for smaller thresholds, and more and longer events for higher thresholds.\newline
All time steps contributing to a running mean below the threshold are considered part of the Dunkelflaute. Therefore, the minimum length of a Dunkelflaute is \SI{48}{\hour}.\newline
We categorise Dunkelflauten according to the dominant weather regime in this period. In case several weather regimes contribute the same number of hours to a Dunkelflaute, the Dunkelflaute is associated with the weather regime that occurs at the onset of the Dunkelflaute since this is the weather regime that triggers the Dunkelflaute in the first place. With this definition, there are eight categories of Dunkelflauten: Dunkelflauten related to one of the seven weather regimes or to the ``no regime''.\newline
As will be shown later (Figure\,\ref{fig:DistributionDF}), Dunkelflauten occur mainly in winter. To capture the sequence of challenging renewable supply conditions, we consider years centred on winter, starting in July and ending in June of the following year. For simplicity, we refer to these periods as extended winter.

\par\null

\subsection{Electricity demand}\label{chap:electricitydemand}
Dunkelflauten events not only put stress on the energy system from the supply side, due to low wind and solar power availability, but also from the demand side, as persistent cold temperatures raise the electricity demand for building heating. Therefore, we also incorporate electricity demand data from \citet{Bloomfield2020} into our analysis, which was built using a multiple-linear regression model for 28 European countries from 1979 to 2018. The model focuses on weather-dependent parameters, therefore neglecting human behavioural factors (day-of-week and long-term socio-economic trends). The demand time series we use can be interpreted as the demand which would have been expected on each weather-day in 1979--2018, with no day-of-week effects and the prevailing socio-economic conditions of 2017. This data reflects the temperature dependence of the electricity demand in 2017. Due to electrification in the space heating sector in Germany \citep{BDEW2022}, it can be expected that the electricity demand will become more temperature-dependent in future scenarios, so results with historical demand model are likely to underestimate the significance of the weather-dependent demand. Data is limited to daily, weather-dependent demand, excluding the diurnal cycle of demand \citep{Bloomfield2020}.

\par\null

\section{Results}\label{Results}
This section first gives a general overview of Dunkelflauten in Germany and their link to weather regimes (Section\,\ref{chap:characteristics}). Subsequently, differences in the origin of Dunkelflauten are revealed by analysing composites of different atmospheric field variables (Section\,\ref{chap:metvariables}). In Section\,\ref{chap:lc_df_characteristics}, we investigate whether weather regime life cycles associated with Dunkelflauten differ from general regime life cycle characteristics. In Section\,\ref{chap:demandinDF} we consider the electricity demand model of \citet{Bloomfield2020} to relate our results from Section\,\ref{chap:metvariables} to the electricity demand during Dunkelflauten events.

\par\null

\subsection{Characteristic of Dunkelflauten in Germany}\label{chap:characteristics}
By our definition, a Dunkelflaute is a rare event indicating a period of low combined wind and solar power output. In the 40\,year period from 1979--2018, we detect 169\,Dunkelflauten which cumulates to an average of four Dunkelflauten per year.
\begin{figure*}[h!]
\includegraphics[width=1.0\columnwidth]{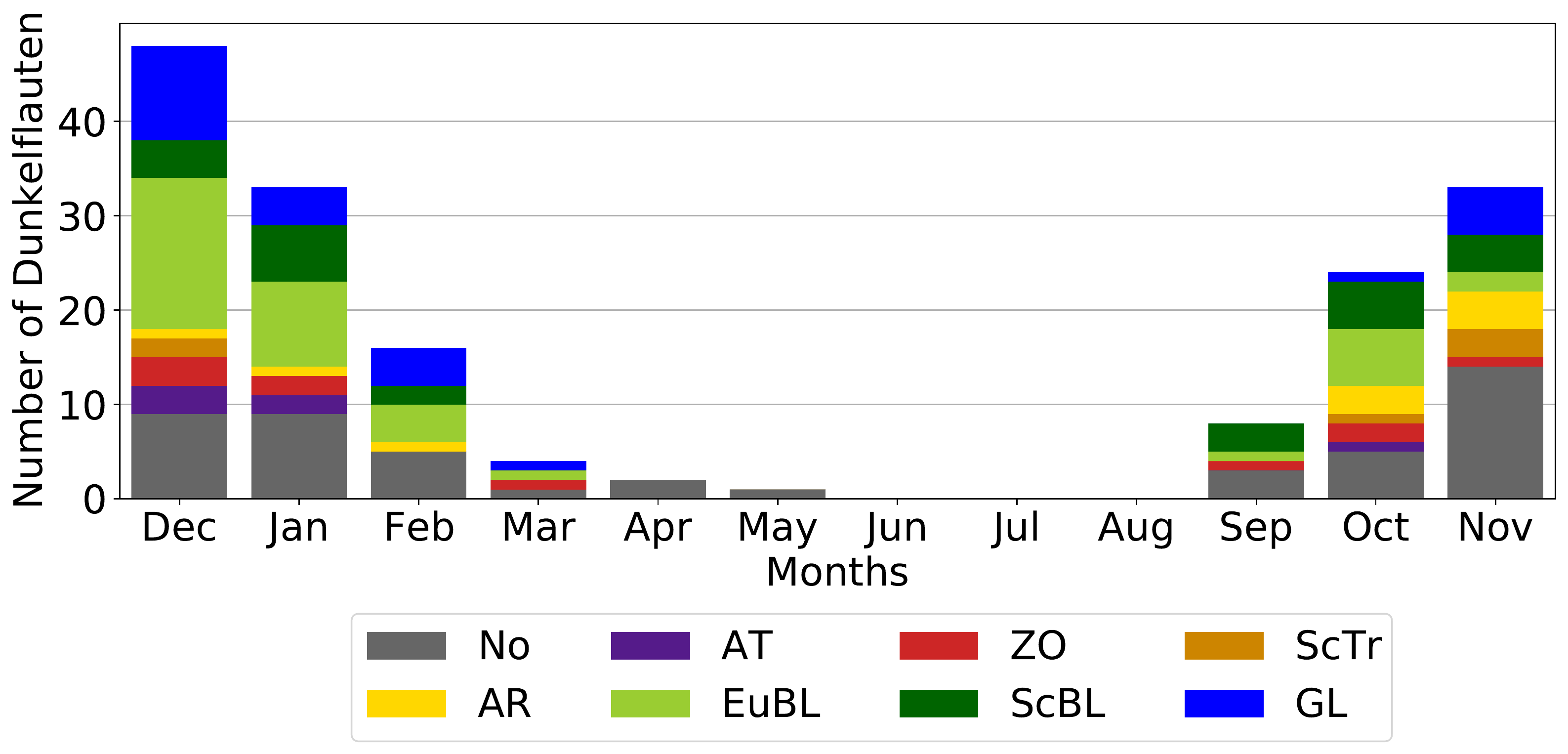}
\caption{Monthly distribution of Dunkelflauten with the dominant weather regimes of Dunkelflauten indicated by the colour of the bars. The amount of Dunkelflauten is measured on an absolute scale for the 40\,year period from 1979--2018.}
\label{fig:DistributionDF}
\end{figure*}
These Dunkelflauten have a pronounced seasonal cycle (Figure\,\ref{fig:DistributionDF}) with the bulk of Dunkelflauten in autumn (SON) and winter (DJF) and no Dunkelflaute in summer (JJA). The seasonality of Dunkelflauten is explainable by the seasonal cycle of the capacity factors (Figure\,\ref{fig:capacityfactor}).\newline
The distribution width of the combined capacity factor, using the 48\,hour running mean in the 40\,year period, is the smallest in summer, and the minimal value of the summer distribution (0.064) is above the Dunkelflauten threshold. In the winter half-year, the combined capacity factor distribution varies on a larger range, occasionally reaching values below the Dunkelflauten threshold. Values below the threshold are due to the combination of low 48\,hour running mean solar capacity factors (average of 0.053 in winter time) and occasional drops in the wind capacity factors below the threshold. In winter, solar power cannot always compensate for the occasional lack of wind power. In summer, the solar capacity factor (average of 0.139) is well above the Dunkelflauten threshold and can thus compensate for periods of low wind capacity factors.\newline
The frequency of Dunkelflauten does not only vary throughout the year but also interannually. On average, 4.3\,Dunkelflauten occur per year (Supplementary material\,\ref{fig:DistributionDF_year}). The most Dunkelflauten per extended winter period have occurred in 1995/1996 and 1996/1997 with 9 and 10\,Dunkelflauten respectively. The average length of a Dunkelflaute is 3.4\,days (82\,hours) with a maximum length of 8\,days (Supplementary material\,\ref{fig:DFlengthdistribution}). The shortest possible length of a Dunkelflaute is set to 2\,days by definition. The length of Dunkelflauten has a similar seasonal cycle as the amount of Dunkelflauten, with the longest Dunkelflauten in December. Supplementary material\,\ref{fig:calendar_plot} provides a detailed visualisation of all Dunkelflauten over the years.
\begin{figure}[h!]
\includegraphics[width=1.0\columnwidth]{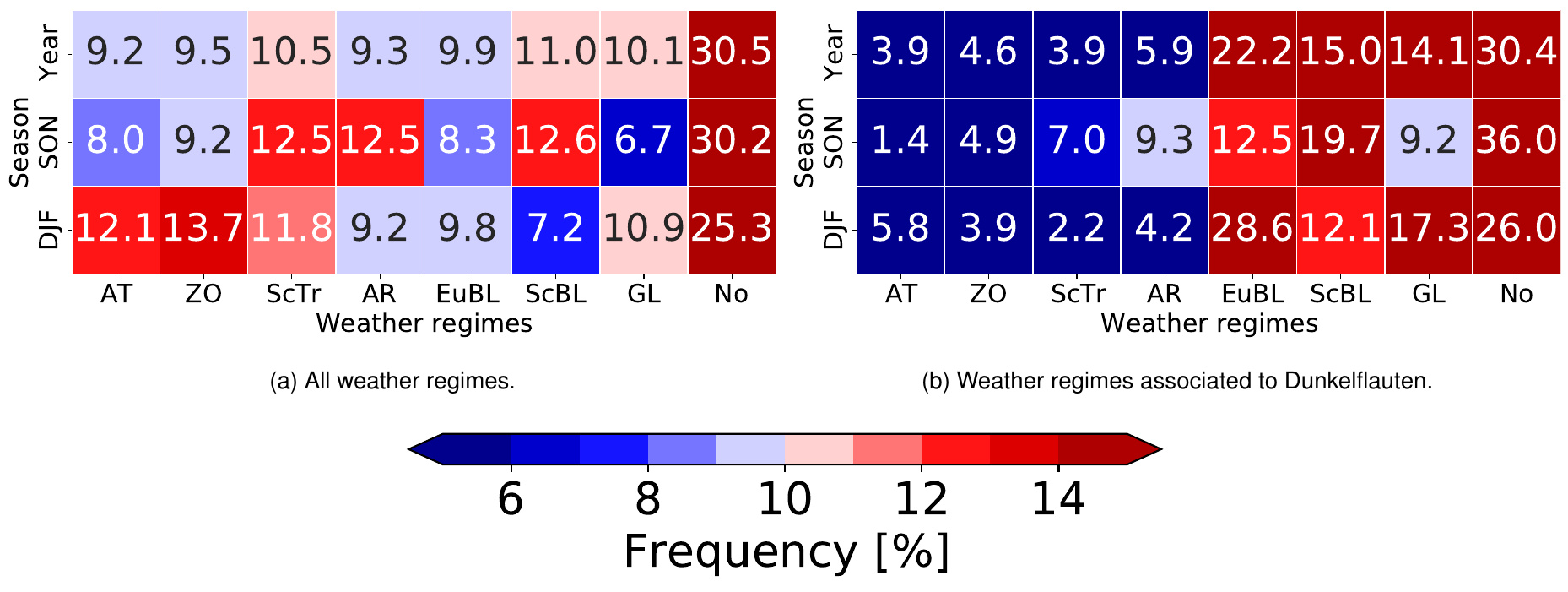}
\caption{Weather regime frequencies for all weather regimes (a) and all weather regimes associated with a Dunkelflaute (b). Shown for all seasons of the year combined, only autumn (SON) and only winter (DJF). Data for the years 1979 to 2018.}
\label{fig:WRfrequency}
\end{figure}\newline
With the help of weather regimes for the Northern Atlantic-European region, we can describe and forecast large-scale circulation, and wind and solar irradiation patterns. The frequency distribution of weather regimes during Dunkelflauten differs from its climatological distribution (Figure\,\ref{fig:WRfrequency}).\newline
The year-round climatological frequency distribution of weather regimes is well balanced throughout cyclonic (Atlantic Trough, Zonal Regime, Scandinavian Trough) and blocked weather regimes (Atlantic Ridge, European Blocking, Scandinavian Blocking, Greenland Blocking) with frequencies between 9.2\% and 11.0\% (Figure\,\ref{fig:WRfrequency}a, top row). However, for Dunkelflauten periods the distribution of the weather regimes is not in balance (Figure\,\ref{fig:WRfrequency}b, top row). Dunkelflauten occur predominantly in three of the four blocked weather regimes, European, Scandinavian and Greenland Blocking, among which Dunkelflauten occur most frequently during European Blocking (22.2\%). As Dunkelflauten mainly occur in autumn and winter (Figure\,\ref{fig:DistributionDF}), we compare the frequency distributions for these seasons separately (Figure\,\ref{fig:WRfrequency}a and\,\ref{fig:WRfrequency}b, second and third row). The preference of Dunkelflauten occurrence during the three blocked weather regimes over the cyclonic weather regimes remains but differences in the preferred frequency of blocked regimes occur. In autumn, Scandinavian Blocking is by 7\% more frequent during Dunkelflauten compared to all autumn days (12.6\% vs. 19.7\%, Figure\,\ref{fig:WRfrequency}). In winter, the frequency of European and Greenland Blocking is increased by 19 and 6\% (increase from 9.8\% up to 28.6\% and from 10.9\% up to 17.3\%, respectively, Figure\,\ref{fig:WRfrequency}) during Dunkelflauten compared to the overall frequency in winter. Thus, Dunkelflauten preferentially occur during the blocked European, Scandinavian and Greendland Blocking regimes.

\par\null

\subsection{Meteorological parameters during Dunkelflauten}\label{chap:metvariables}
\begin{figure*}[h!]
\includegraphics[width=1.0\columnwidth]{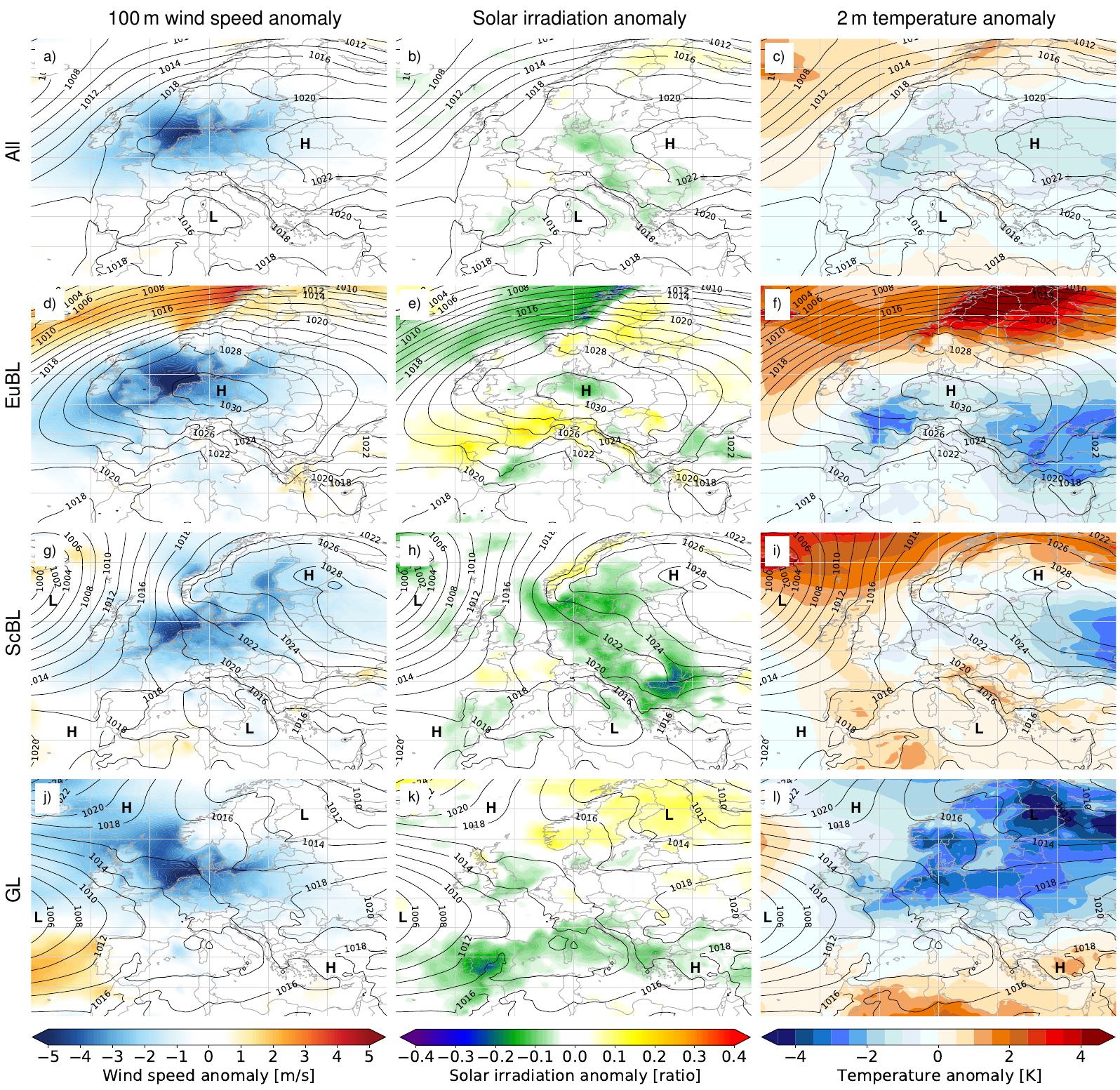}
\caption{Composites of atmospheric field variables for different types of Dunkelflauten. The rows show all Dunkelflauten events (a-c), European Blocking Dunkelflauten (d-f), Scandinavian Blocking Dunkelflauten (g-i) and Greenland Blocking Dunkelflauten (j-l). The first column shows \SI{100}{\meter} wind speed anomalies (with respect to a 30\,day running mean climatology) as shading, the second column the anomaly in the daily fraction of maximum solar insolation (with respect to the 30\,day running mean climatology) and the third column \SI{2}{\meter} temperature anomalies (with respect to the 30\,day running mean climatology). Each composite shows the absolute sea level pressure with a \SI{2}{\hecto\pascal} contour interval.}
\label{fig:AtmosphericComposites}
\end{figure*}

To further shed light on the meteorological conditions during Dunkelflauten and to estimate the severity of Dunkelflauten during different regimes in terms of surface weather, we analyse anomaly composites (respective to the 30\,day running mean climatology) of different atmospheric field variables (\SI{100}{\meter} wind speed, solar irradiation, \SI{2}{\meter} temperature, sea level pressure) in the North-Atlantic-European region. The solar irradiation is given as the ratio of the daily sums of the surface net solar radiation (SSR in ERA5 reanalysis) and the surface net solar radiation for clear skies (SSRC in ERA5 reanalysis). This quantity reflects the fraction of the maximum possible daily insolation \citep[cf.][]{Grams2017}. \newline
The composites of all Dunkelflauten events (Figure\,\ref{fig:AtmosphericComposites}a--c) represent the mean atmospheric conditions.
Low wind speeds are associated with weak surface pressure gradients. Weak pressure gradients can occur in the centre of a high-pressure system or the saddle point between multiple high and low-pressure systems. On average, weak pressure gradients in the centre and northwestern edge of a surface anticyclone over Europe cause low wind speeds during Dunkelflauten in Germany (Figure\,\ref{fig:AtmosphericComposites}a). Insolation is only marginally altered in comparison to the 30\,day running mean climatology (Figure\,\ref{fig:AtmosphericComposites}b). This suggests that Dunkelflauten are primarily due to a lack of wind power. On average, \SI{2}{\meter} temperature anomalies of up to \SI{2}{\kelvin} occur in Central Europe (Figure\,\ref{fig:AtmosphericComposites}c).\newline
Analysing Dunkelflauten separated by weather regimes demonstrates that the conditions for Dunkelflauten can be achieved by different atmospheric conditions (Figure\,\ref{fig:AtmosphericComposites}d--l). Thus, the mean conditions for all events do not reflect variability in meteorological conditions during Dunkelflauten events imposed by different weather regimes.\newline
The three dominant Dunkelflauten types, European, Scandinavian and Greenland Blocking (shown in Figure\,\ref{fig:WRfrequency}), all indicate low wind speeds in the North Sea region due to weak pressure gradients but the pressure system patterns differ between the Dunkelflauten types (Figure\,\ref{fig:AtmosphericComposites}d,g,j). For the three dominant Dunkelflauten types, there are only weak modulations of irradiation (Figure\,\ref{fig:AtmosphericComposites}e,h,k). Strong differences in the temperature anomalies amongst these Dunkelflauten types exist (Figure\,\ref{fig:AtmosphericComposites}f,i,l).\newline
During European Blocking Dunkelflauten low wind speeds in Northern Germany and the North Sea region occur in the centre of a high-pressure system centred over Germany (Figure\,\ref{fig:AtmosphericComposites}d).\newline
In contrast, low wind speeds for Scandinavian and Greenland Blocking Dunkelflauten occur in a region of weak pressure gradients due to a quadripole of two high and two low-pressure systems centred over Germany. For Scandinavian Blocking Dunkelflauten (Figure\,\ref{fig:AtmosphericComposites}g), the low pressure dominates near Iceland and in the Mediterranean, high pressure prevails in the vicinity of the the Azores and Scandinavia. The quadripole is reversed for Greenland Blocking Dunkelflauten (Figure\,\ref{fig:AtmosphericComposites}j). Low pressure centres are located over the Atlantic north of the Azores and Scandinavia, high pressure extends over the Icelandic region and Southeastern Europe.\newline
Solar irradiation is only marginally altered for the different Dunkelflauten types. Minor positive (negative) anomalies occur in the south (north) of Germany during European Blocking Dunkelflauten (Figure\,\ref{fig:AtmosphericComposites}e). For Scandinavian Blocking Dunkelflauten, only a negative solar irradiation anomaly is detected in northern Germany (Figure\,\ref{fig:AtmosphericComposites}h), and no solar irradiation anomalies in Germany are present during Greenland Blocking Dunkelflauten (Figure\,\ref{fig:AtmosphericComposites}k).\newline
For European Blocking Dunkelflauten, the \SI{2}{\meter} temperature, in particular in Germany but also in Western and Eastern Europe in general, is up to \SI{2}{\kelvin} below the 30\,day running mean climatology (Figure\,\ref{fig:AtmosphericComposites}f). For Scandinavian Blocking Dunkelflauten, the south of Germany experiences marginally warmer and the north of Germany marginally colder \SI{2}{\meter} temperatures in comparison to climatology (Figure\,\ref{fig:AtmosphericComposites}i). During Greenland Blocking Dunkelflauten, Germany is under the influence of substantial negative \SI{2}{\meter} temperature anomalies, which are up to \SI{4}{\kelvin} colder compared to the 30\,day running mean climatology (Figure\,\ref{fig:AtmosphericComposites}l). The negative \SI{2}{\meter} temperature anomaly is present in Western, Central, Eastern, and Northern Europe, as well as in Russia.\newline
Analysing the temperature anomalies in more detail for Greenland Blocking Dunkelflauten shows that the negative anomalies in Germany and Northern Europe are already present 6\,days prior to the onset of the Dunkelflauten (Supplementary material\,\ref{fig:GLT2MAextended}). Cold polar air is advected to Europe prior to the onset of the Dunkelflaute. Subsequently, the cold air mass becomes stationary in Germany and even endures the Dunkelflaute itself. The long-lasting negative temperature anomalies amplify the potential stress that an energy system faces during and after Dunkelflauten related to Greenland Blocking.\newline
In summary, exploring the meteorological conditions during Dunkelflauten related to different weather regimes unveiled that not all Dunkelflauten are caused by high pressure over Germany and go along with cold conditions. For Dunkelflauten related to Greenland Blocking, Germany is located in a saddle point between weather systems. Importantly, Greenland Blocking Dunkelflauten are cold Dunkelflauten, causing unusually cold conditions for a prolonged period, and, therefore, likely stress the energy system.

\par\null

\subsection{Characteristics of weather regimes associated to Dunkelflauten}\label{chap:lc_df_characteristics}
The strong link of Dunkelflauten to specific weather regimes raises the question of whether regime life cycles associated with Dunkelflauten differ from the general characteristics of regime life cycles. Therefore, we  now explore life cycle characteristics such as the regime duration and the relation of Dunkelflauten occurrence and regime onset and decay.
\begin{figure}[h!]
\medskip
\centering
\includegraphics[width=0.7\columnwidth]{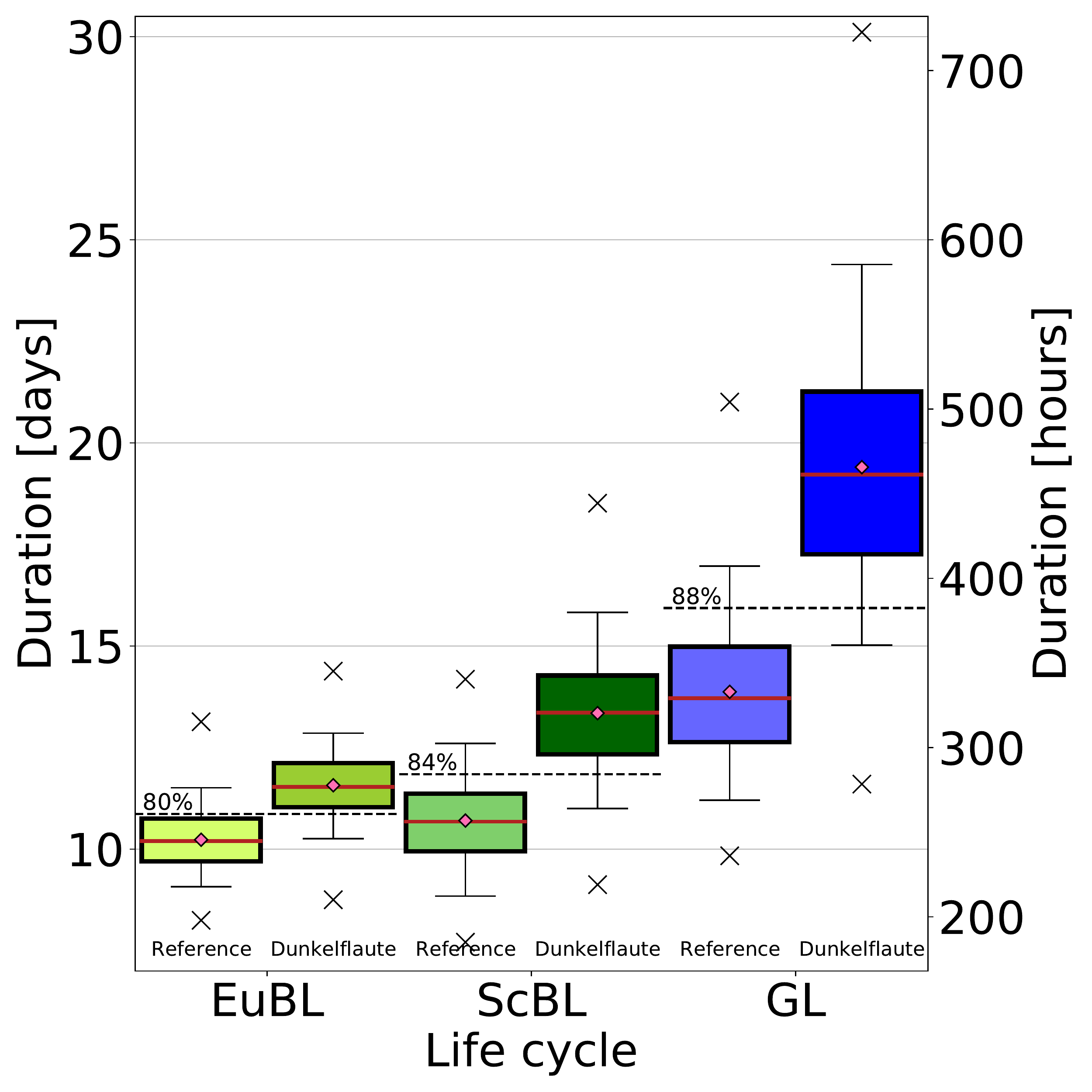}
\caption{Comparison of the mean regime life cycle duratiplotsons of life cycles not necessarily associated with Dunkelflauten (reference sample, left columns) and life cycles associated with Dunkelflauten (Dunkelflauten sample, right columns). Compared are Box and Whisker plots of the distributions for the European, Scandinavian and Greenland Blocking weather regimes. The horizontal dashed lines indicate the confidence intervals between the different lengths of the two samples. Pink diamonds represent the mean, red horizontal lines the median, boxes the 25--75\% percentile, whiskers the 5--95\% percentiles and outliers the minimum and maximum values.}
\label{fig:lifecycle_length}
\end{figure}\newline
First we investigate the duration of weather regimes with Dunkelflauten. Due to the small sample size of weather regimes, we do not simply compare the average life cycle lengths but instead perform a bootstrapping: we create two samples for each weather regime category from which we randomly draw sets of regime life cycles with the same number of elements. One sample, the Dunkelflauten sample, includes only life cycles associated with Dunkelflauten. The other sample, the reference sample, includes all life cycles in the 40 year period occurring in a $\pm$45\ day window around the day of the year of the Dunkelflauten constraints. By limiting the reference time frame, we ensure a fair comparison that retains the seasonal cycle of Dunkelflauten events. From both samples we draw 1000 random sets with repetition and the same number of elements as Dunkelflauten events. For each of the 1000 random sets we then calculate the mean life cycle length to obtain a distribution of the mean life cycle length for both samples. The significance of the life cycle length differences between the two samples is expressed by the significance level, the percentile at which both samples have the same life cycle length. Results are more significant if the significance level is closer to 100\%.\newline
The mean life cycle length distribution in the reference sample reveals differences between the European, Scandinavian and Greenland Blocking weather regimes (Figure\,\ref{fig:lifecycle_length}, left columns). Whereas European and Scandinavian Blocking weather regime life cycles are active for 10.2 and 10.7\,days respectively, Greenland Blocking life cycles have a mean lifespan of 13.8\,days.\newline
Comparing the Dunkelflauten sample (Figure\,\ref{fig:lifecycle_length}, right columns) to the reference sample (Figure\,\ref{fig:lifecycle_length}, left columns) for each weather regime separately shows significantly longer life cycle lengths for the Dunkelflauten samples. Life cycles associated with European and Scandinavian Blocking Dunkelflauten have a mean length of 11.6 and 13.3\,days respectively (an increase of 14 and 24\% concerning the reference sample). These longer life cycle lengths are significant on the 80\% and 84\% level compared to the reference samples. For life cycles associated with Greenland Blocking Dunkelflauten, the difference is more pronounced with an average life cycle length of 19.4\,days (an increase of 41\%) and an 88\% significance level.\newline
The results of this bootstrapping method indicate that Dunkelflauten occur in longer-lived life cycles compared to life cycles in the same seasonal period.
\begin{figure}[h!]
\medskip
\centering
\includegraphics[width=0.7\columnwidth]{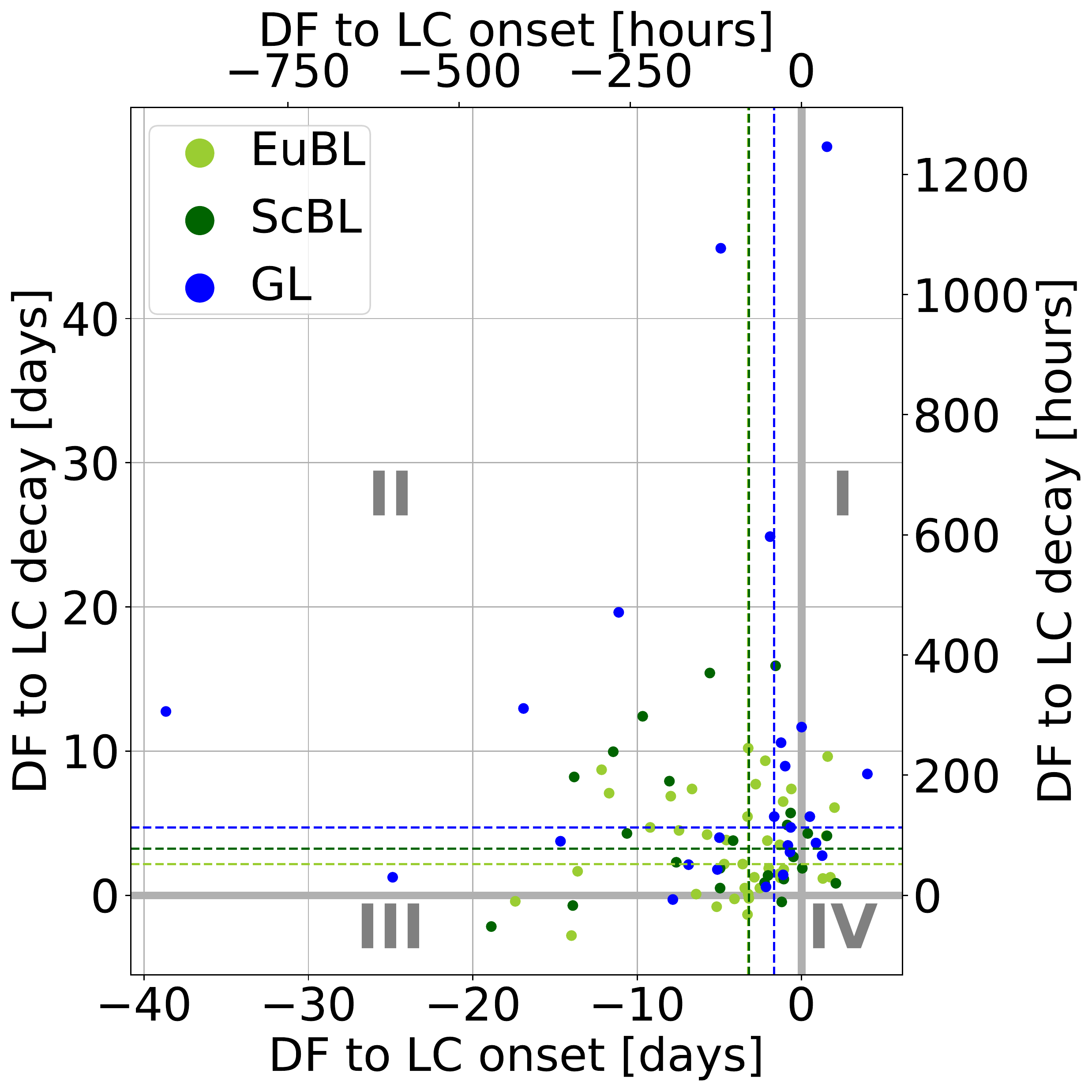}
\caption{Onset and decay of Dunkelflauten (DF) in relation to the onset and decay of the associated regime life cycle (LC) for European, Scandinavian and Greenland Blocking Dunkelflauten. The median time differences between the onsets and decays for each Dunkelflauten category are represented by the dashed vertical and horizontal lines, respectively. The four quadrants distinguish the following categories: I. A Dunkelflaute sets in before the life cycle and decays during the active life cycle; II. A Dunkelflaute sets in during the associated life cycle and decays during the active life cycle; III. A Dunkelflaute sets in during an active life cycle and decays after the decay of the life cycle; IV. A Dunkelflaute sets in prior to the onset of the life cycle and lasts longer than the life cycle; therefore decays after the decay of the life cycle.}
\label{fig:DF_LC_onset_decay}
\end{figure}\newline
Considering the average life cycle length of 11--20\,days and the Dunkelflauten length of 3--9\,days, the subsequent question is whether Dunkelflauten occur at specific times of the associated life cycle.\newline
Dunkelflauten are mostly fully embedded in the regime life cycle associated with the Dunkelflaute (Figure\,\ref{fig:DF_LC_onset_decay} quadrant II, indicating that the Dunkelflauten onset is after the life cycle onset and the Dunkelflauten decay is prior to the life cycle decay). In numbers, this translates to an average (median) onset of Dunkelflauten 3.3/3.2/1.7\,days after the onset of the life cycle and an average decay 2.2/3.2/4.7\,days prior to the decay of the life cycle for European/Scandinavian/Greenland Blocking, respectively. Thus, Dunkelflauten occur mainly in the well-developed and stable phases of regime life cycles.

\par\null

\subsection{Electricity demand during Dunkelflauten}\label{chap:demandinDF}
With negative \SI{2}{\meter} temperature anomalies during European and Greenland Blocking Dunkelflauten (shown in Section\,\ref{chap:metvariables}), we expect a higher heating demand in Germany due to colder than normal temperatures in wintertime. The combination of a Dunkelflauten event with low renewable power output and an increased electricity demand due to cold temperatures is likely to put stress on the energy system.
\begin{figure}[h!]
\medskip
\centering
\includegraphics[width=1.0\columnwidth]{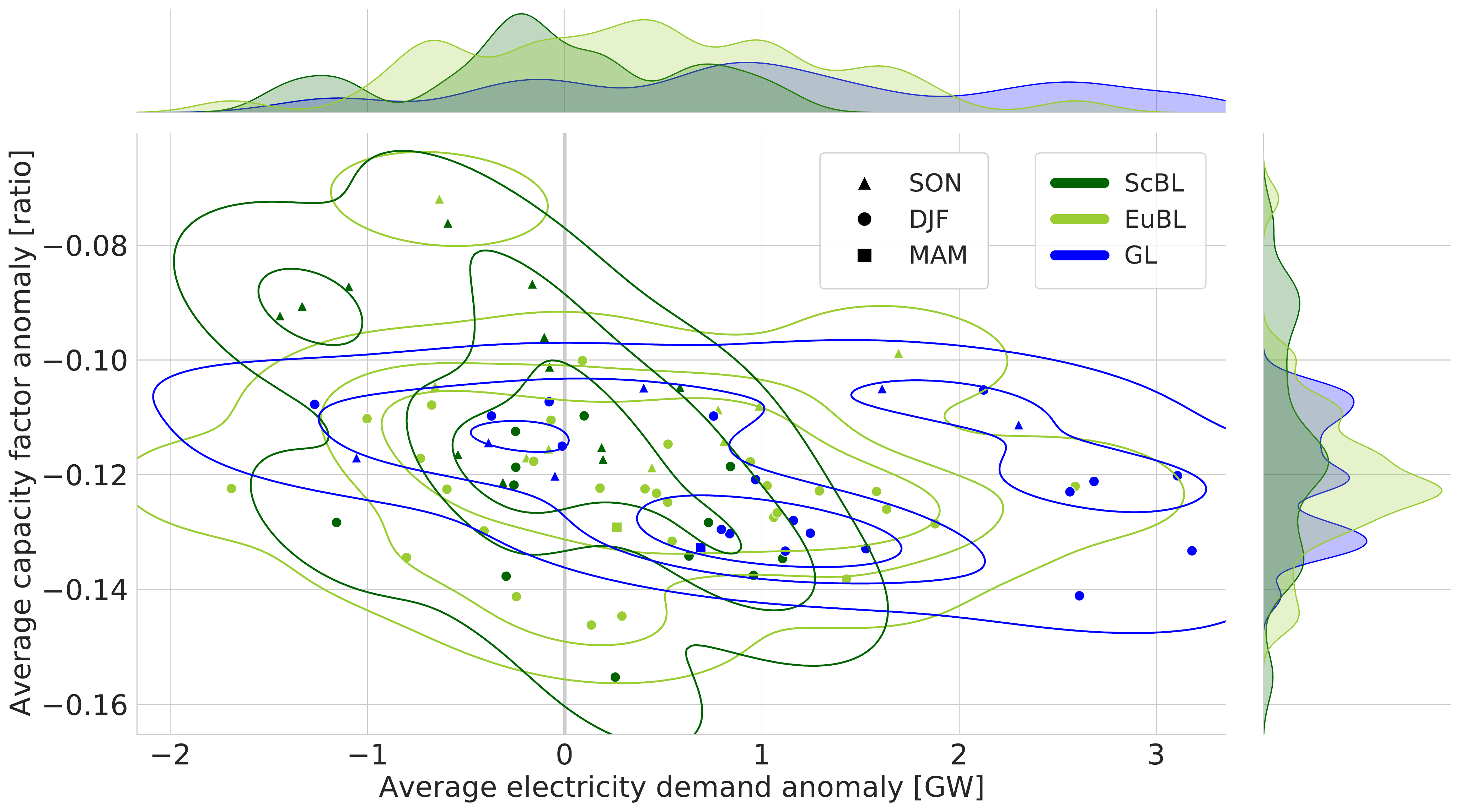}
\caption{Average capacity factor and electricity demand anomalies during different Dunkelflauten types in Germany. The electricity demand is based on the prevailing socioeconomic conditions of 2017. All European, Scandinavian, and Greenland Blocking Dunkelflauten are represented with their regime colour. The shape of the marker indicates the season of each Dunkelflaute. The 2D density distribution of each Dunkelflauten type is given as shading in the main figure and as 1D distribution in the two marginal figures.}
\label{fig:DF_powersystemmodel}
\end{figure}\newline
The capacity factors during the three dominant Dunkelflauten types are reduced on average from 0.18 to 0.06 (Figure\,\ref{fig:DF_powersystemmodel}, right marginal figure), which is a relative reduction of 66\% compared to winter climatology.\newline
Whereas there is no anomalously high average electricity demand for Scandinavian Blocking Dunkelflauten (Figure\,\ref{fig:DF_powersystemmodel}, top marginal figure), the average electricity demand during Dunkelflauten is marginally increased for European Blocking Dunkelflauten (average increase of \SI{0.5}{\giga\watt}) and more strongly increased for Greenland Blocking Dunkelflauten (average increase of \SI{1.0}{\giga\watt}, but values also reaching up to \SI{3.2}{\giga\watt}). Absolute values of the electricity demand in the period from November to March range from \SI{44.5}{\giga\watt} up to \SI{76.5}{\giga\watt}. Therefore, during Dunkelflauten, the electricity demand may rise by up to 7\%, using the socioeconomic conditions of 2017. Although we did not directly investigate the dependence of electricity demand and \SI{2}{\meter} temperature, the latter is likely a main meteorological driver of demand \citep{Bloomfield2020} (Figure\,\ref{fig:DF_powersystemmodel}). Also, as shown in Section\,\ref{chap:metvariables}, marked temperature anomalies occur in Germany during Dunkelflauten. In particular, persistent low temperatures accompany the low power output during the cold Greenland Blocking Dunkelflaute (Figure\,\ref{fig:AtmosphericComposites}l and Supplementary material\,\ref{fig:GLT2MAextended}). Thus we conclude that Greenland Blocking Dunkelflauten are likely the most challenging periods for the operation of an energy system with high shares of renewables.\newline
An even larger increase of the electricity demand can be expected for the future due to the transition to a fully renewable energy system so that this problem amplifies.

\par\null

\section{Discussion and conclusions}\label{Discussion}
The present study links large-scale atmospheric circulation patterns, called weather regimes, and periods with low wind and solar power in Germany, called Dunkelflauten.\newline
Following \citet{Kaspar2019}, Dunkelflauten are defined as periods with a low combined capacity factor of solar PV, onshore wind and offshore wind in Germany, lasting for at least two days, using the Atlite energy system model framework and meteorological data from ERA5 reanalysis. We then link one of seven year-round weather regimes (or ``no regime''; \citet{Grams2017}), to each Dunkelflauten event from 1979–2018.\newline
We find Dunkelflauten in Germany mainly in autumn and winter. This seasonal effect is due to the low mean solar capacity factor in winter. Thus Dunkelflauten are primarily caused by low wind speed conditions.\newline
Differentiating Dunkelflauten by the prevalent large-scale atmospheric flow, using weather regimes, helps to identify different atmospheric patterns leading to Dunkelflauten in Germany. Three blocking weather regimes are most frequently associated with Dunkelflauten: European Blocking with 22\% of all Dunkelflauten, Scandinavian Blocking with 15\%, and Greenland Blocking with 14\%. A high-pressure system over Germany and Central Europe is the most frequent Dunkelflauten pattern and associated with the European Blocking weather regime.\newline
The meteorological conditions during Greenland, Scandinavian, and European Blocking Dunkelflauten differ: A high pressure system extends over Central Europe during European Blocking Dunkelflauten and Germany is located in the centre of the high pressure with prevalent low pressure gradients and consequently low winds. In contrast, during Greenland Blocking and Scandinavian Blocking Dunkelflauten, Central Europe is located in a saddle point of a quadripole of two high and two low-pressure systems, likewise causing low pressure gradients and low winds in Germany. Thus, it is essential to analyse Dunkelflauten not as one composite but separately between different weather regime patterns. Greenland Blocking Dunkelflauten are considered to be cold Dunkelflauten. These Dunkelflauten periods are up to \SI{4}{\celsius} colder than the 30\,day running climatology, increasing the electricity demand from electrical heating. Increased electricity demand in periods of lower than normal power output by renewable energy sources puts stress on the energy system.\newline
Weather regime life cycles associated with Dunkelflauten also differ in their general characteristics: Life cycles associated with Dunkelflauten are longer-lived than usual, especially for Greenland Blocking weather regimes, with life cycles on average 5\,days longer than the November to March (NDJFM) climatology (19 days compared to 14 days). The life cycles of weather regimes begin well before the onset of the Dunkelflauten and end thereafter. Thus Dunkelflauten occur when weather regime life cycles are well-established. This is useful from a forecast perspective, as knowledge about upcoming blocking weather regimes can help prepare the energy system for an imminent supply shortage.\newline
Our results confirm findings from previous studies and extend the knowledge of Dunkelflauten. \citet{Kaspar2019} and \citet{Drucke2021} also find Dunkelflauten in Germany mainly in autumn and winter. They categorise Dunkelflauten by Grosswetterlagen (Grosswetterlagen classify the circulation in Europe with 29\,different weather types focusing more on the regional conditions than the continental-scale large-scale weather regimes used here) and identify the ``Grosswetterlage GWL9'' as the most frequent pattern for Dunkelflauten in Germany. GWL9 has comparable characteristics to the European Blocking weather regime. \citet{VanderWiel2019b} identify the NAO-- weather regime (using the four weather regime classification by \citet{Vautard1990, Michelangeli1995, Cassou2008}), which strongly correlates with the Greenland Blocking weather regime, to be the scenario with reduced energy production and increased energy demand for Europe. The increased electricity demand in Greenland Blocking Dunkelflauten, using the electricity demand model of \citet{Bloomfield2020}, is likely to intensify in the future due Germany's transition to electrical heating. In the past 10\,years, the contribution of electrical heating to the heating structure of new buildings has doubled, from 24\% in 2012 up to 50\% in 2022 \citep{BDEW2022}. \citet{Otero2022} show the link of energy compound events in Germany, simultaneous episodes of low renewable energy production (wind plus solar power) and high electricity demand, and weather regimes. Energy compound events in Germany are more frequent in European and Greenland Blocking weather regimes. These results are consistent with the cold Dunkelflauten in our research.\newline
We extend the knowledge of Dunkelflauten in Germany by analysing weather regime life cycles and meteorological conditions associated with different Dunkelflauten. Dunkelflauten, especially the cold Greenland Blocking Dunkelflauten, are positioned in the well-established phase of longer-lived weather regime life cycles. Furthermore, Dunkelflauten not only occur when high pressure prevails but also in conditions when Germany is in the middle between pronounced weather systems elsewhere.\newline
The link of Dunkelflauten to weather regimes provides a forecast opportunity on the subseasonal to seasonal (S2S) range. \citet{Bloomfield2021b} showed that pattern-based methods, such as weather regimes, outperform grid-point forecasts for lead times larger than 12\,days for the European national power forecasts. Weather regime forecasts can fill the gap between short-range weather prediction and long-range seasonal outlooks for the energy sector \citep{White2017}. With the results of our study, we expect to more accurately forecast Dunkelflauten using weather regimes, as the likelihood of a potential Dunkelflaute could be identified not only by the forecasted weather regime but also by the forecasted duration of a weather regime life cycle.\newline
Based on our results in combination with the results by \citet{Bueler2021}, showing the modulation of surface weather by weather regimes and the promising skill of weather regimes on the subseasonal to seasonal range, we see weather regime forecasts as an essential tool for energy system operators to prepare for multi-day supply shortages.

\section*{Author contributions}

\textbf{Fabian Mockert}: Conceptualization (equal); methodology (lead); software (lead); formal analysis (lead); data curation (lead); writing -- original draft (lead); writing -- review and editing (equal); visualization (lead). 
\textbf{Christian Grams}: Conceptualization (equal); writing -- review and editing (equal); funding acquisition (equal); project administration (equal); supervision (equal).
\textbf{Tom Brown}: Conceptualization (equal); writing -- review and editing (equal); funding acquisition (equal); project administration (equal); supervision (equal).
\textbf{Fabian Neumann}: Conceptualization (equal); methodology (supporting); software (supporting); data curation (supporting); writing -- review and editing (equal); supervision (equal).

\section*{Acknowledgements}\label{acknowledgements}

This project was initiated and partially funded through a KIT Young Investigator Network (YIN) Grant 2019. FM further acknowledges from the KIT Center for Mathematics in Sciences, Engineering and Economics under the seed funding programme. The contribution of CMG is funded by the Helmholtz Association as part of the Young Investigator Group ``Sub-seasonal Predictability: Understanding the Role of Diabatic Outflow'' (SPREADOUT, grant VH-NG-1243). The research was partially embedded in the subprojects A8 of the Transregional Collaborative Research Center SFB/TRR 165 ``Waves to Weather'' (https://www.wavestoweather.de) funded by the German Research Foundation (DFG). FN and TB gratefully acknowledge funding from the Helmholtz Association under grant VH-NG-1352. We thank Hannah Bloomfield for providing the demand time series and the members of the Energy Meteorology and Dynamical Processes groups at the University of Reading for the fruitful discussions.

\section*{Conflict of interest}
All authors declare that they have no conflicts of interest.

\bibliography{bibliography/library_DF.bib%
}
\newpage
\renewcommand{\thesection}{S}
\section{Supplementary material}
\setcounter{figure}{0}
\renewcommand\thefigure{\thesection\arabic{figure}}
\begin{figure}[h!]
\includegraphics[width=1.0\columnwidth]{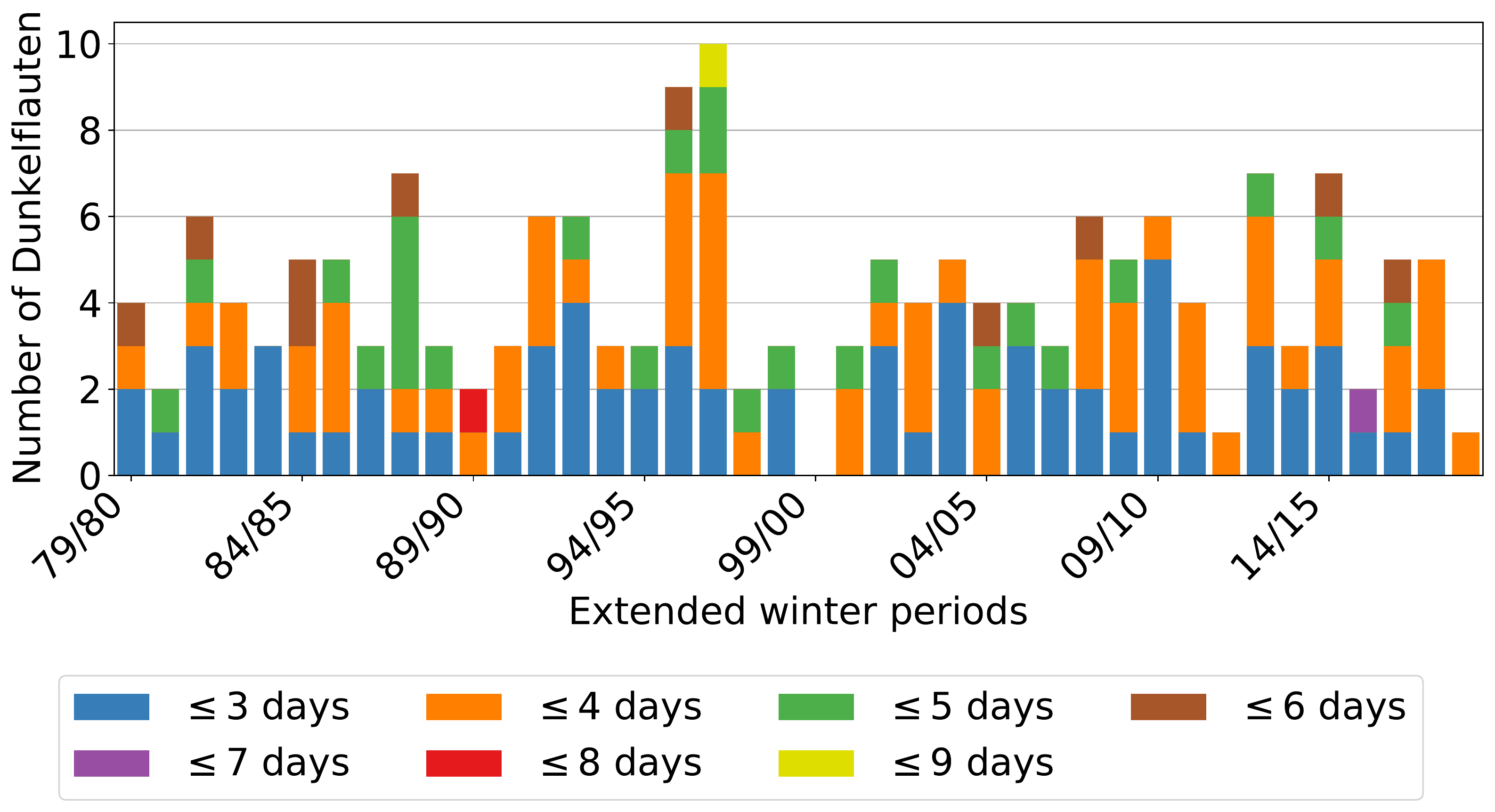}
\caption{Amount and length of Dunkelflauten over the 40\,year period from 1979--2018. The years are represented as extended winter periods, starting in July of the first year and ending in June of the second year (x-axis).}
\label{fig:DistributionDF_year}
\end{figure}

\begin{figure}[h!]
\medskip
\includegraphics[width=1.0\columnwidth]{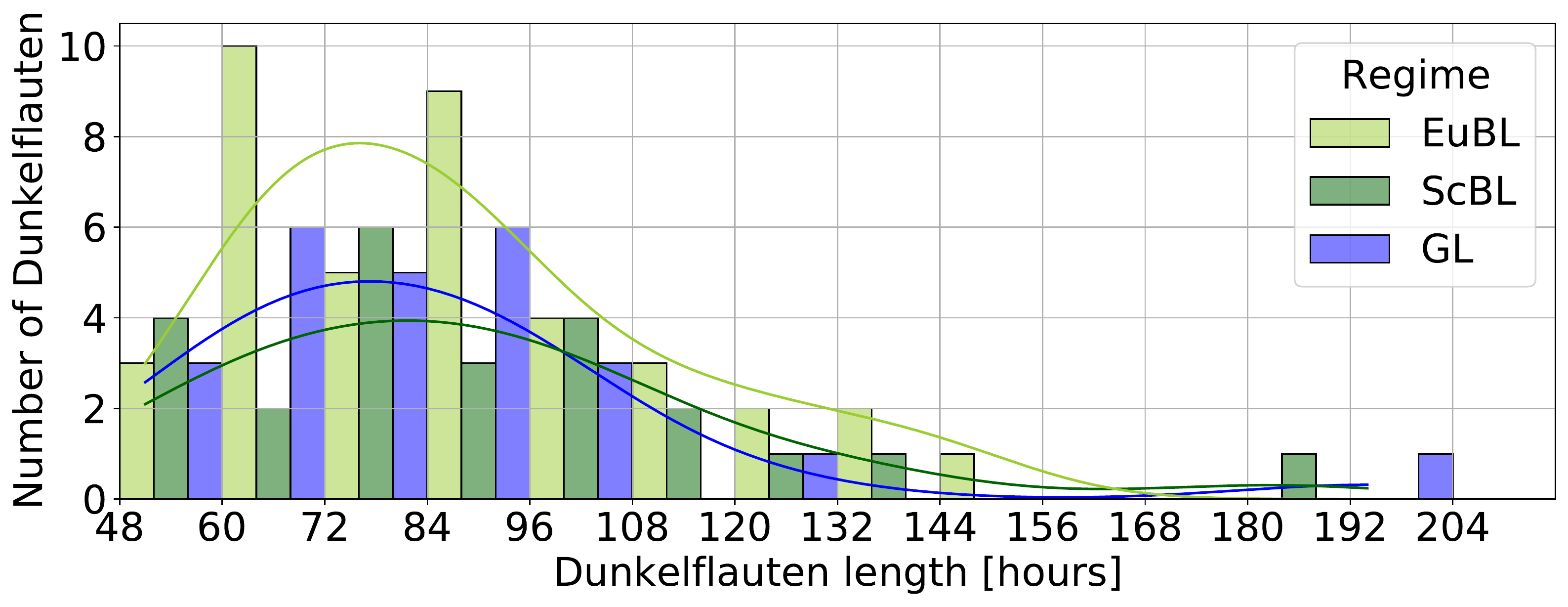}
\caption{Length distribution of Dunkelflauten for the European, Scandinavian and Greenland Blocking weather regimes.}
\label{fig:DFlengthdistribution}
\end{figure}

\begin{figure*}[h!]
\medskip
\centering
\includegraphics[width=1.0\columnwidth]{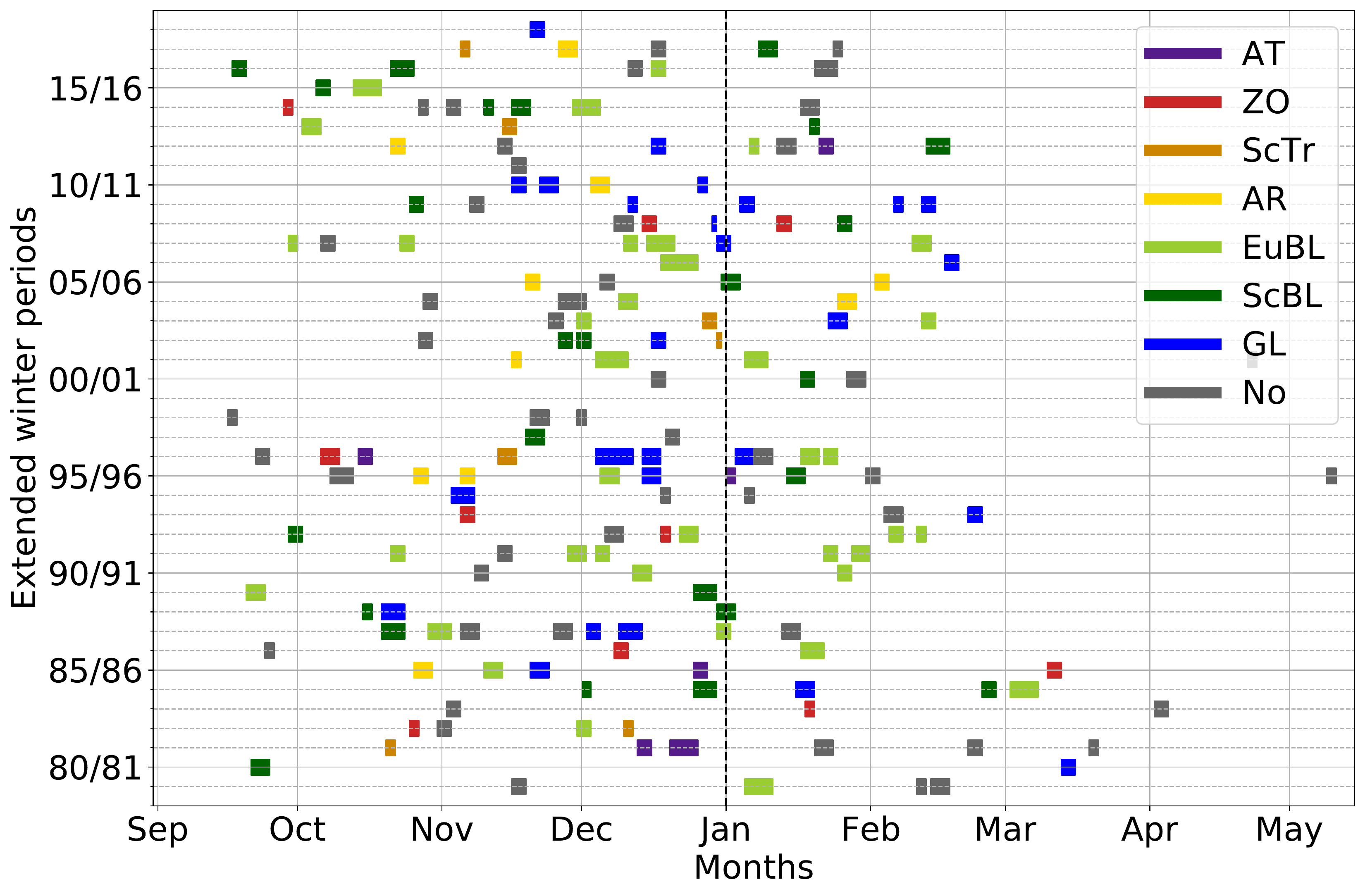}
\caption{Calendar plot of all Dunkelflauten in the 40\,year period from 1979--2018. The colours indicate the main weather regimes occurring in the Dunkelflauten phase. The years are represented as extended winter periods (y-axis) and months are shown from September to May (x-axis).}
\label{fig:calendar_plot}
\end{figure*}

\begin{figure*}[h!]
\includegraphics[width=1.0\columnwidth]{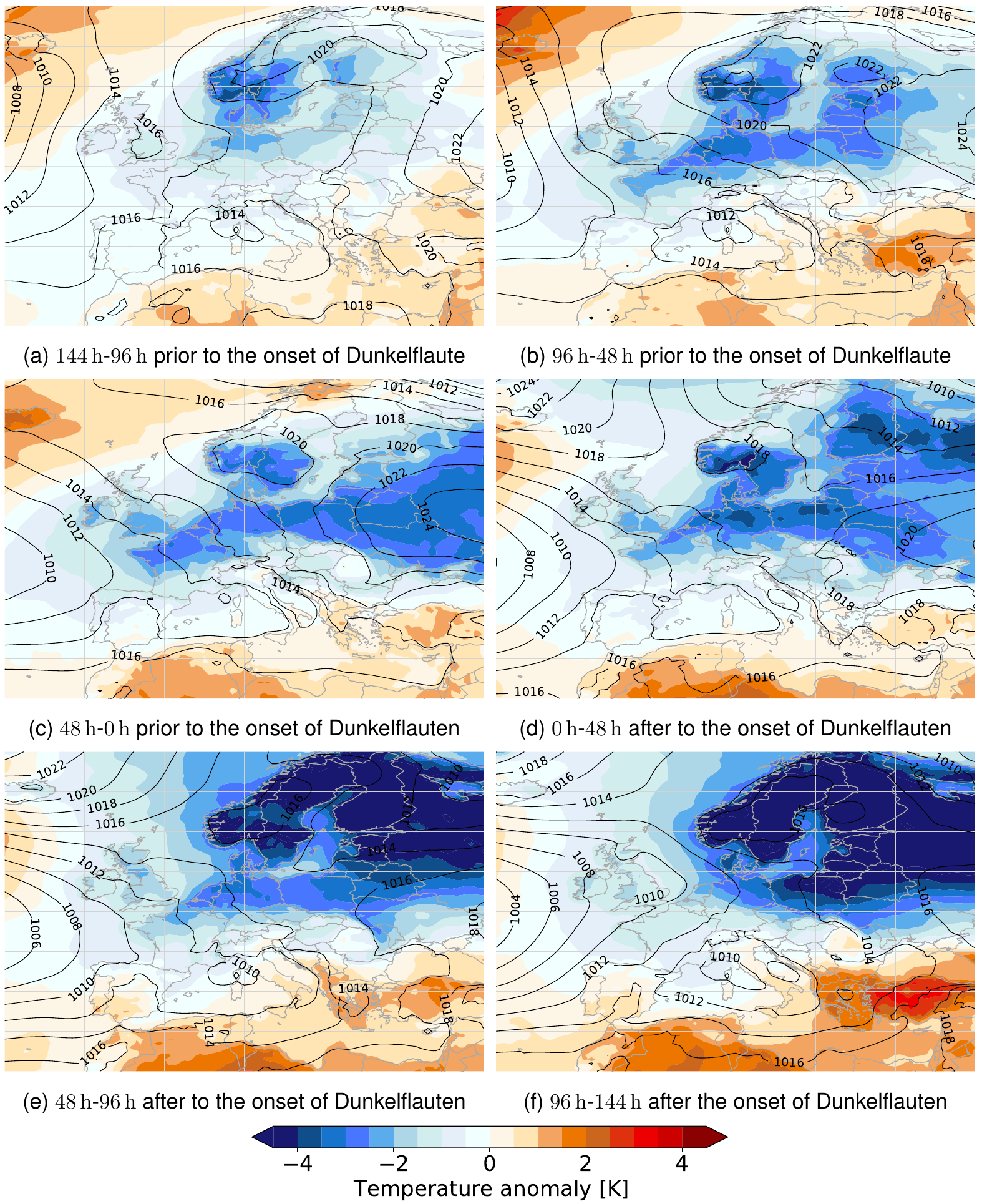}
\caption{Mean temporal evolution of sea level pressure (contouring every \SI{2}{\hecto\pascal}) and \SI{2}{\meter}\,temperature anomalies (shading every \SI{0.5}{\kelvin}) in the phase building up to Greenland Blocking Dunkelflauten (a-c, starting 6\,days prior to the onset of Dunkelflauten) and the phase during and after Greenland Blocking Dunkelflauten (d-f, ending 6\,days after the onset of Dunkelflauten).}
\label{fig:GLT2MAextended}
\end{figure*}
\end{document}